\documentstyle[11pt,epsfig]{article}
\newcommand{\zp}[3]{Z.\ Phys.\ {\bf C#1} (19#2) #3}
\newcommand{\pl}[3]{Phys.\ Lett.\ {\bf B#1} (19#2) #3}
\newcommand{\np}[3]{Nucl.\ Phys.\ {\bf B#1} (19#2) #3}
\newcommand{\prd}[3]{Phys.\ Rev.\ {\bf D#1} (19#2) #3}

\def\simgt{\rlap{\lower 3.5 pt \hbox{$\mathchar \sim$}} \raise 1pt \hbox {$>$}}
\def\simlt{\rlap{\lower 3.5 pt \hbox{$\mathchar \sim$}} \raise 1pt \hbox {$<$}}

\newcommand{\beq}{\begin{equation}}
\newcommand{\eeq}{\end{equation}}
\newcommand{\bea}{\begin{eqnarray}}
\newcommand{\eea}{\end{eqnarray}}

\catcode`\@=11
\def\section{\@startsection{section}{1}{\z@}{3.5ex plus 1ex minus .2ex}
{2.3ex plus .2ex}{\large\bf}}
\def\thesection{\arabic{section}.}

\def\appendix{\setcounter{section}{0}
 \def\thesection{Appendix \Alph{section}:}
 \def\theequation{\Alph{section}.\arabic{equation}}}

\def\@citex[#1]#2{\if@filesw\immediate\write\@auxout{\string\citation{#2}}\fi
  \def\@citea{}\@cite{\@for\@citeb:=#2\do
    {\@citea\def\@citea{,\penalty\@m}\@ifundefined
       {b@\@citeb}{{\bf ?}\@warning
       {Citation `\@citeb' on page \thepage \space undefined}}%
\hbox{\csname b@\@citeb\endcsname}}}{#1}}

\def\citer{\@ifnextchar [{\@tempswatrue\@citexr}{\@tempswafalse\@citexr[]}}
\def\@citexr[#1]#2{\if@filesw\immediate\write\@auxout{\string\citation{#2}}\fi
  \def\@citea{}\@cite{\@for\@citeb:=#2\do
    {\@citea\def\@citea{--\penalty\@m}\@ifundefined
       {b@\@citeb}{{\bf ?}\@warning
       {Citation `\@citeb' on page \thepage \space undefined}}%
\hbox{\csname b@\@citeb\endcsname}}}{#1}}
\relax

\begin{document}

\thispagestyle{empty}
\begin{flushright}
\hfill{CERN-TH/95-324}\\
\hfill{DESY 95-240}
\end{flushright}
\vskip 3cm
\begin{center}{{\bf\Large HEAVY-QUARK PRODUCTION IN $e\gamma$ SCATTERING}}
\vglue 1.8cm
\vglue 0.6cm
\begin{sc}
 Eric Laenen\\
\vglue 0.15cm
\end{sc}
{\it CERN TH-Division\\
1211-CH, Geneva 23, Switzerland}
\vglue 0.25cm
and\\
\vglue 0.5cm
\begin{sc}
Stephan Riemersma\\
\end{sc}
\vglue 0.15cm
{\it
DESY - Zeuthen\\
Platanenallee 6, D-15735 Zeuthen, Germany}
\vglue 0.5cm
\end{center}

\vglue 2cm
\begin{abstract}
\par \vskip .1in \noindent
Open heavy flavour production at $e^+e^-$ colliders in deeply inelastic
$e\gamma$ scattering has an interesting feature:
the structure function $F_2(x,Q^2)$ for this process is
calculable for $x>0.01$, and is essentially proportional
to the gluon density in the photon for smaller $x$
values. We give estimates for event rates at LEP2 and a Next
Linear Collider
in $x,Q^2$ bins, and present differential
distributions in the transverse momentum and rapidity of the heavy quark
for the case of charm.
We include all next-to-leading-order
QCD corrections, and find theoretical uncertainties
are well under control.

\end{abstract}

\vfill
\begin{flushleft}
CERN-TH/95-324\\
DESY 95-240\\
December 1995
\end{flushleft}

\newpage

\setcounter{page}{1}

\noindent 

\section{Introduction}

A copious amount of heavy quarks produced in two-photon collisions will
be generated at high energy $e^+e^-$ colliders such as LEP2 and a
future 
Next Linear Collider (NLC). The largest fraction of these will come 
from so-called no-tag events, in which neither of the leptons 
generating the photons is seen.
However, due to the high energy and luminosity of these machines, 
a sizeable number of them will be produced in events where either the
outgoing $e^+$ or $e^-$ is tagged (``single-tag events'').
In such events, the photon coming from the tagged lepton is far
off-shell 
and spacelike, hence this reaction amounts to deeply inelastic
electroproduction 
of heavy quarks on a photon target. An important aspect of on-shell
photons initiating a hard scattering is that they may behave either as
pointlike particles (``direct'' process), or as hadronic (vector)
particles
(``resolved'' process) \cite{plhad}. 
In no-tag events, both the direct and resolved
process contribute to the cross section.
At LEP2 they contribute in about equal amounts 
to the cross section for charm production in two-photon collisions
\cite{DKZZ}.
In such events one can separate these components
by making use of forward detectors and 
using the presence of a remnant jet of spectator partons in resolved processes
as a separator \cite{k0}. 
The interesting feature of single-tag heavy-quark events is that
this separation occurs quite naturally in the deeply inelastic Bjorken 
scaling variable $x$, as we will show below. Moreover, the presence of the
heavy quark mass ensures that this separation is unambiguous to
next-to-leading order (NLO) in QCD.
The direct process is directly calculable in QCD and is free from 
such phenomenological inputs as parton densities in the photon.
The resolved channel on the other hand is directly proportional to
the poorly known gluon density in the photon.
Therefore a reasonable sample of single-tag heavy-quark events allows
one to confront simultaneously a well-controlled perturbative QCD calculation 
with experiment, and constrain the gluon density in the photon.

In the past (open) heavy quark (mainly charm) production 
in two-photon collisions at $e^+e^-$ colliders has been
difficult to observe in experiments due to the low charm acceptance.
The reaction $e^+e^- \rightarrow e^+e^- D^{*\pm} X$ has been
thoroughly studied experimentally.
The existence of the $D^{*\pm}$ has been inferred either
from direct reconstruction 
\cite{jade,tpcgg,dM} or from unfolding the distribution of soft pions 
\cite{topaz2,sp} resulting from its decay. In addition, 
measurements have been made using soft leptons \cite{sl,topazsl} 
and kaons \cite{k0} to tag the charm.
JADE \cite{jade} and TPC/Two-Gamma~\cite{tpcgg} have performed
experimental studies of the reaction 
$e^+e^- \rightarrow e^+e^- D^{*\pm} X$ with one outgoing
lepton tagged at low average value $\langle Q^2
\rangle$ of the  
momentum transfer squared of the tagged
lepton (below $1 \,\,({\rm GeV}/c)^2$).
TOPAZ~\cite{topaz2} has performed a study as well at somewhat larger
$\langle Q^2 \rangle$. 
The total number of events obtained was however very small
(about 30 for TOPAZ). See \cite{LEP2rep} for a more extensive
review of the experimental situation.

The cross section and 
single particle distributions for $\gamma\gamma\rightarrow c\bar c$ have
been calculated to next-to-leading order (NLO) in QCD
in \cite{DKZZ}, and agree with experimental
results. Correlations in the direct channel were studied in \cite{kl}. 
For the single-tag case, the structure
functions $F_2^{\gamma}$ and $F_L^{\gamma}$ for charm
production were calculated to NLO in QCD in \cite{LRSN}.
In this letter we employ the results of \cite{LRSN} to estimate the expected
number of single-tag events in $x,Q^2$ bins for LEP2 and a NLC, and examine 
differential distributions in the transverse momentum and
rapidity of the charm quark.

In section 2 we describe the formalism and explain
our notation. In section 3 we show 
the structure function $F_2^{\gamma}$ for charm production,
give estimates of event rates in $x,Q^2$ bins for the case of 
LEP2 and a NLC and show single-charm-quark differential
distributions.  We summarize and conclude in section 4.

\section{Formalism}

We consider the reaction (see Fig.1)
\begin{equation}
e^-(p_e) + e^+ \rightarrow e^-(p_e') + e^+ + 
Q(p_1)+X\,,\label{one}
\end{equation}
where 
$Q(p_1)$ is a heavy quark with momentum $p_1$ and 
$X$ denotes any hadronic state 
allowed by quantum-number 
conservation laws.
\begin{figure}[hbtp]
\vglue 7cm
\vbox{\includegraphics{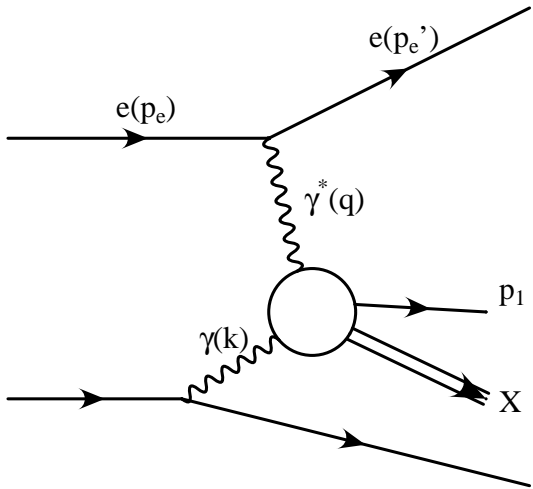} }
\caption{\it Charm production in a single-tag two-photon event.}
\end{figure} 
When the outgoing electron is tagged,
this reaction is dominated by the subprocess
\begin{eqnarray}
\gamma^*(q) + \gamma(k) \rightarrow Q(p_1) +X\,,\label{two}
\end{eqnarray}
where one of the photons is highly virtual and the other
is almost on-mass-shell and transversely polarized. 
The case where the positron is 
tagged is completely equivalent. This process is described by the
differential cross section
\begin{eqnarray}
\label{three}
\frac{d^4\sigma}{dxdQ^2dT_1 dU_1} 
&=& \int dz \,  f_\gamma^e (z, \frac{S}{m_e^2})
\,  \frac{2\pi\alpha^2}{x\,Q^4} \\
&\times& \left[ ( 1 + (1-y)^2 )
\frac{d^2 F^\gamma_2(x,Q^2,m^2)}{dT_1 dU_1} -y^2 
\frac{d^2 F^\gamma_L(x,Q^2,m^2)}{dT_1 dU_1} \right]\,,
\nonumber \end{eqnarray}
where the $d^2F^\gamma_k(x,Q^2,m^2)/dT_1 dU_1$ $(k=2,L)$ denote the 
(doubly differential) deeply inelastic
photon structure functions, $\alpha = e^2/4\pi$
is the fine structure constant, and $m$ the heavy quark mass.
$S$ is the center of mass energy squared of the $e^+ e^-$ system.
The Bjorken scaling variables $x$ and $y$ are
defined by
\begin{equation}
x = \frac{Q^2}{2k\cdot q} \,, \qquad y = \frac{k\cdot q}{k \cdot p_e}
\,, \qquad q = p_e - p_e'\,. \label{four}
\end{equation}
Further, we define the variables
\begin{equation}
T_1 = T-m^2 = (k-p_1)^2-m^2 \;,\;\;
U_1 = U-m^2 = (q-p_1)^2-m^2\, .
\end{equation}
The momenta of the off-shell photon and the on-shell photon
obey the relations $q^2 = -Q^2 <0$
and $k^2 \approx 0$ respectively. Because the photon with momentum $k$ is 
almost on-mass-shell, 
eq. (3) 
is written in the Weizs\"acker-Williams
approximation: the function $f^e_\gamma(z,S/m_e^2)$ represents
the probability of finding a photon $\gamma(k)$ in the positron,
with longitudinal momentum fraction $z$. It is given by \cite{FMNR}
\begin{equation}
  f^e_{\gamma}(z) = \frac{\alpha}{2\pi} \left\{
  \frac{1+(1-z)^2}{z}\, \ln \frac{k^2_{\rm max}}{k^2_{\rm min}}
  - 2 m_e^2 z(\frac{1}{k^2_{\rm min}}
 -\frac{1}{k^2_{\rm max}})\right\}\, ,
\label{fww}
\end{equation}
where $k^2_{\rm min}  = (z^2 m_e^2)/(1-z)$ and $k^2_{\rm max}  = 
 (1-z) \left(E_b \theta_{\rm max}\right)^2+ (z^2 m_e^2)/(1-z)$.
Here $E_b=\sqrt{S}/2$ 
is the lepton beam energy and $\theta_{\rm max}$ is the anti-tag
\footnote{The \rm maximum angle below which no energy deposit
in forward calorimeters may be seen, thus
selecting events in which the lepton generating the target
photon goes down the beampipe.}
angle.
In the rest of the calculation $k^2=0$.

In 
eq.~(3)
the differential structure functions 
can be represented as
\begin{eqnarray}
dF_k^{\gamma}(x,Q^2,m^2) &= &
dC_{k,\gamma}(x,\frac{Q^2}{m^2})+
\sum_{i=q,\bar q, g}\int_x^{z_{\rm max}} \frac{dz}{z}  
f_i^{\gamma}(\frac{x}{z},\mu_f^2)\, dC_{k,i}(z,\frac{Q^2}{\mu_f^2},m^2) 
\nonumber \\
& = & dF_k^{\gamma,PL} + dF_k^{\gamma,HAD}\,, \label{seven}
\end{eqnarray}
where $d$ represents $d^2/dT_1 dU_1$ and 
$z_{\rm max} = Q^2/(Q^2+4m^2)$. The two terms represent
the contributions to the structure functions from 
pointlike (or ``direct'') and hadronic (or ``resolved'') photons,
respectively.
The $f_i^{\gamma}$ are photonic parton densities and
the $dC_{k,i}\,(i=q,\bar q,g,\gamma)$ are 
(differential) Wilson coefficient functions, and 
$\mu_f$ is the mass factorization scale.
At lowest order, only the photonic gluon density
appears in eq.~(\ref{seven}).
$F_2^{\gamma}$ and $F_L^{\gamma}$ were calculated 
in \cite{LRSN} to NLO
in QCD by computing all $O(\alpha_s)$ corrections
to the coefficient functions $C_{k,i}$. We use the 
results of this calculation for the present analysis.
Coupling constant renormalization was performed in the 
$\overline{\rm MS}$ scheme, modified such that heavy flavours
decouple in loops when small momenta flow into the fermion loop, 
and mass renormalization in
the on-shell scheme. Mass factorization was also
performed in the $\overline{\rm MS}$ scheme. See
\cite{LRSN} for further details.
Note the absence of the scale $\mu_f$ in the second
term on the right hand side of eq.~(\ref{seven}). This is due to the fact that,
through NLO, the heavy quark mass prevents a collinear singularity
from occurring in the pointlike piece. For massless
quarks this singularity is subtracted at scale $\mu_f$
and absorbed into the hadronic piece.
As a consequence $F_k^{\gamma,PL}$ is calculable --
the only theoretical uncertainties stemming from
$\alpha_s$ and the heavy quark mass-- whereas 
$F_k^{\gamma,HAD}$ is primarily sensitive to the gluon density in the
photon, in analogy to the proton
case.
We will see that the two components on the right hand side
in eq.~(\ref{seven}) dominate
in different regions of the Bjorken variable $x$.
>From 
eq. (3)
we will estimate the numbers
of events expected per $x,Q^2$ bin at LEP2 and a future NLC.

In analyzing charm production, we will also show differential
distributions  
in transverse momentum and rapidity of the detected charm quark
in the $\gamma^* e^+$ c.m. frame.
These quantities are derived from the variables defined
in the above as follows.
We define $W$ to be the invariant mass squared
\begin{equation}
W = (q+k)^2
\end{equation}
and $W'= W+Q^2$. The transverse mass of the heavy quark
in the $\gamma^* \gamma$ frame is determined by
\begin{equation}
W'^2 m_T^2 = W' T_1 U_1 + Q^2 T_1^2 + Q^2 W' T_1 \, .
\end{equation}
The transverse mass is the same in the $\gamma^* e^+$
frame. Further, in the $\gamma^* \gamma$ frame
the energy of the heavy quark is 
\begin{equation}
\tilde{E} = \frac{-Q^2-T_1-U_1}{2\sqrt{W}}\, .
\end{equation}
Its longitudinal momentum can be found from the 
condition $\tilde{P}_L^2 = \tilde{E}^2-m_T^2$, so its
rapidity in the $\gamma^* \gamma$ frame is 
\begin{equation}
\tilde{y} = \frac{1}{2}
\ln \left(\frac{\tilde{E}+\tilde{P}_L}{\tilde{E}-\tilde{P}_L}\right)\, .
\end{equation}
In the $\gamma^* e^+$ frame, 
$y = \tilde{y} + \frac{1}{2}\ln(z)$
(we define a particle to have positive rapidity when it travels
into the hemisphere from which the $\gamma^*$ originates).

\section{Results}

\begin{figure}[hbtp]
\vglue 8.5cm
\vbox{\includegraphics{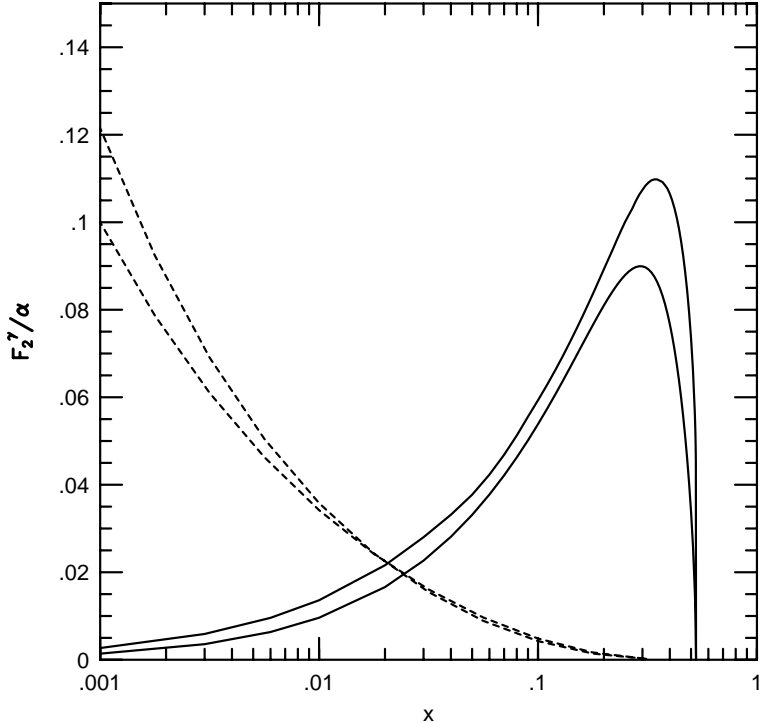} }
\caption{\it
Hadronic (dashed lines) and pointlike (solid lines) components
of $F_2^{\gamma}/\alpha$ vs. $x$ at $Q^2= 10\,({\rm GeV}/c)^2$.
The lower solid line is the LO, the upper one the NLO case.
The lower dashed line at $x=0.001$ is the NLO, the upper one the
LO case.}
\end{figure} 

We first list our choices
for various quantities used in the calculations of this
section. Deviations from them will be
explicitly indicated.
We used for $m$, the heavy quark mass, the value
$1.5$ GeV for charm 
and $4.75$ GeV for bottom. We assumed
a center of mass energy $\sqrt{S} = 180$ (500) GeV for LEP2 (NLC).
For the strong coupling we took
a one-loop running $\alpha_s$ with three (four) light flavours and
$\Lambda= 232$ (200) MeV at LO and a two-loop running  $\alpha_s$ with
$\Lambda= 248$ (200) MeV at NLO for charm (bottom).
We put the renormalization scale equal to the factorization
$\mu_f$ and chose $\mu_f=Q$.
We also imposed the following requirements on the angle
and energy of the tagged lepton: 
$\theta_{\rm tag} > 30\,(40) \, {\rm mrad}$ 
for LEP2 (NLC) and $E_{\rm tag} > E_{\rm beam}/2$, where $E_{\rm beam} = \sqrt{S}/2$.
These cuts imply a minimum $Q^2$ of 3.65 ${\rm GeV^2}$ for LEP2 and 50
${\rm GeV^2}$ at our hypothetical NLC.
Furthermore we used the GRV \cite{GRV} parton densities
for the hadronic photon results, and the Weizs\"acker-Williams density of 
\cite{FMNR} with a
$\theta_{\rm max} = \theta_{\rm tag}$ for the equivalent
target photon spectrum. 
%
%
For the total integrated luminosity of LEP2 (NLC) we used 
$\int {\cal L} dt = 500\, {\rm pb}^{-1}$ ($20 \, {\rm fb}^{-1}$).
At a NLC, beamstrahlung effects are expected to play an important r\^{o}le.
We include them here by adopting for its spectrum the expression given
in \cite{Chen}, with parameters
$\Upsilon_{eff} = 0.039$ and $\sigma_z=0.5$ mm \cite{Schulte},
corresponding to the TESLA design.
When showing differential distributions we will show them
from the point of view of the $\gamma^*e^+$ cm frame, {\it i.e.}
we assume here that the electron is tagged and take the case where the
positron is tagged into account 
by multiplying the results with a factor 2.

We begin by showing in 
Fig.2 the quantities 
$F_2^{\gamma,PL}$ and $F_2^{\gamma,HAD}$
versus $x$ both at LO and NLO, for $Q^2= 10\,({\rm GeV}/c)^2$.
The $O(\alpha_s)$ corrections to both quantities are  
for the most part fairly small in the $x$-region shown. We may
therefore assume that even 
higher order contributions are negligible,
and these components are thus calculated with some accuracy.

The prominent feature in these figures is clearly
the separation in $x$ of the components. 
This fact was already noted in \cite{LRSN}, and we
emphasize it here.
As a result, the opportunity to confront a precise calculation and
measure the small-$x$ photonic gluon density presents itself in one
experiment. 
As mentioned earlier, the difficulty of
efficient charm tagging makes such an experimental study very difficult 
in practice.  To judge
the feasibility of such a study we have integrated
eq.~(3)
for various $x,Q^2$ bins, and obtained
estimates for the number of charm quarks per bin produced
at LEP2 and NLC by both hadronic and pointlike photons.
Note that these estimates use the NLO calculation.

To perform the integrals over $x$ and $Q^2$,
we used fitted versions of the {\em integrated} NLO coefficient functions
$C_{k,i}\,(i=q,\bar q,g,\gamma)$ in eq.~(\ref{seven}), 
as the actual expressions
in \cite{LRSN} are too long for fast evaluation.
By adapting the fitted coefficient functions of
electroproduction of heavy quarks on a proton
target \cite{RSN} to the photon target, we were able to speed up our
computer program by as much as a factor of fifty. 

The results for LEP2 are shown in Tables 1 and 2.
We give in these tables the expected number of events due
to pointlike (Table 1) and hadronic (Table 2) photons.
Assuming a charm tagging efficiency of 1-2\%, we that on average a few tens of 
events per bin should be observable for larger $x$
values. 
Furthermore, to test the stability of the results
we varied the renormalization/factorization scale $\mu_f$
from $Q/2$ to $2Q$. 
The results change relatively little under these variations,
as seen in Tables 1 and 2.
For these LEP2 conditions, we varied in addition the charm quark mass
from $1.3$ GeV to $1.7$ GeV and found that the pointlike contribution
(Table 1) changed under this variation in the charm mass by
less than 10\% in all bins, except in the $x$ bins 
$0.32<x<1$. In these bins, the mass variation causes a variation in
the position of the charm quark production threshold, and from
the shape of $F_2^{\gamma,PL}$ in Fig.2 one may readily 
understand that this causes a large change in the number of events, 
as much as 70\% in the lowest $Q^2$ bin. 
The sensitivity of the hadronic contribution (Table 2) to these changes
is much more uniform in $x$, varying from 30-40\% in the lowest $Q^2$
bin to 10-20\% in the higher $Q^2$ bins.

In Table 3 we give similar estimates for the NLC, both for charm
and bottom production. Here we have summed the contributions
due to pointlike and hadronic photons for each entry. We do
however show separately the number of events due to 
beamstrahlung and Weizs\"{a}cker-Williams bremsstrahlung contributions. 
Note that the number of charmed events here is truly large, even
at large $Q^2$. The number of events containing bottom quarks
is much smaller due to charge and phase space suppression.

Clearly the production rate of charm quarks produced by hadronic
photons is too small at LEP2 to be useful for a good measurement
of the gluon density in the photon. At a NLC the
rates are significantly higher however. We therefore give
in Table 4 the production rate for a NLC in $x,Q^2$ bins for various
choices of parton densities, charm quark mass values and
mass factorization scales.  Examining the composition of these results,
we find very little contribution due to the hadronic channel
from beamstrahlung photons (about 1-2\%).
We further find the contributions due to pointlike photons
both of WW and beamstrahlung origin to be similarly small.
There is therefore a good possibility of measuring the gluon density in the photon
at a NLC, poor tagging efficiency notwithstanding.  The
most significant uncertainty is related to the charm quark mass,
however a differentiation of the gluon densities of GRV \cite{GRV} and
ACFGP \cite{acfgp} seems certainly feasible.

While bottom quark production lends itself to considerably better
tagging efficiencies, the production rate is severely suppressed
relative to charm.
An analysis similar to the one for charm reveals that a measurement
of the photonic gluon density from bottom production 
in $e\gamma$ scattering does not seem possible at a NLC.

We turn to single particle inclusive differential distributions 
in transverse momentum ($p_t$) and rapidity ($y$) of the heavy quark
in the $\gamma^* e^+$ c.m. frame. Due to the low rate of bottom quark
production 
we only show these distributions for the
case of charm. We integrate $x$ and $Q^2$ over the intervals
$ 0.001 < x < 1$ and $3.2 < Q^2 < 320\, (3200)$ GeV$^2$ for 
LEP2 (NLC).  To approximate
detector limitations, we use the restrictions 
$p_t>0.5$ GeV and $|y| < 2$, 
in addition to the cuts listed in the beginning of this section.
The $p_t$ restriction is of course quite loose, but it allows us
to study the behavior of the $p_t$ spectrum at low $p_t$.
Furthermore, for the results shown in Figs.3 and 4 we took the 
factorization (=renormalization) scale  $\mu_f=\sqrt{Q^2+p_t^2}$.

\begin{figure}[hbtp]
\vglue 8.5cm
\vbox{\includegraphics{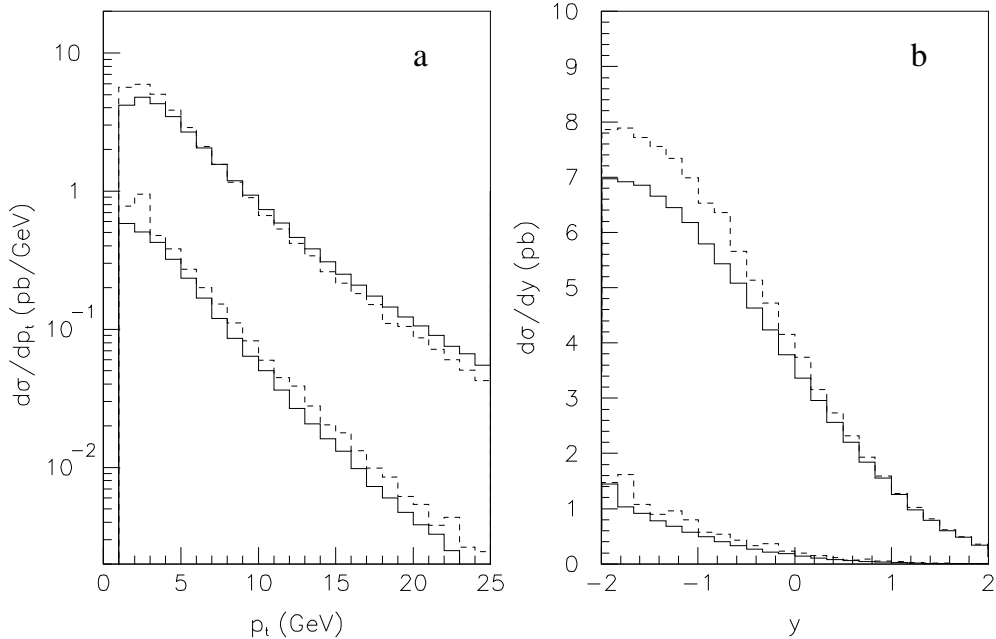} }
\caption{\it
Hadronic and pointlike distributions in $p_t$ and $y$ of the charm quark
in LO and NLO at LEP2. 
a) The upper set of histograms represents 
the pointlike $p_t$-distribution, 
the lower the hadronic. The solid histograms represent the LO case, 
the dashed the NLO. b) Same as in a) but now for the $y$-distribution.}
\end{figure} 

In Fig.3a we present for LEP2
the LO and NLO $p_t$-distributions of the charm quark 
for the pointlike and hadronic piece
separately. In neither instance do the LO and NLO curves
differ significantly. While the hadronic piece
at NLO is larger than the LO at large $p_t$ (which is similar to what
was found for a proton target in \cite{LRSN2}), the reverse is true
for the pointlike piece. The bulk of the events has $p_t < 5$ GeV.
In Fig.3b the same is shown but now for the rapidity distribution.
Clearly the charm quark has predominantly negative rapidity.

\begin{figure}[hbtp]
\vglue 8cm
\vbox{\includegraphics{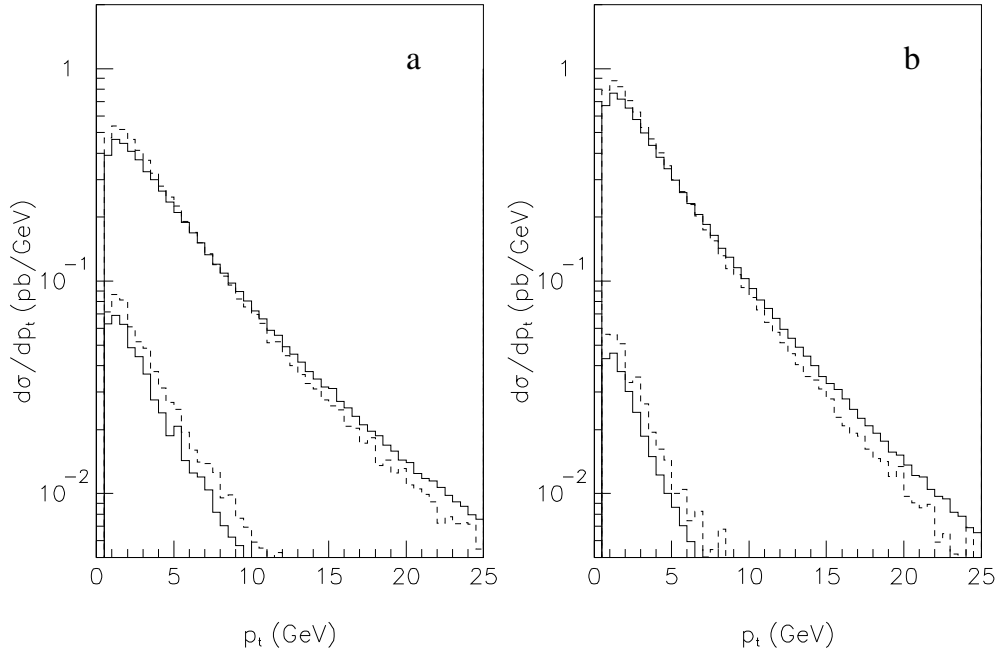} }
\caption{\it
Hadronic and pointlike distributions in $p_t$ of the charm quark
in NLO for the Weizs\"{a}cker-Williams (WW) and beamstrahlung cases 
at a NLC.
a) WW case. The upper set of histograms is the pointlike compoment,
the lower the hadronic. The solid histograms represent the LO
case, the dashed ones the NLO case. b) Same as in a) but now for the 
(TESLA) beamstrahlung spectrum.}
\end{figure} 

In Figs.4a and 4b we present for the NLC
the $p_t$ spectrum results due to pointlike
and hadronic photons for a 
Weizs\"{a}cker-Williams spectrum and a beamstrahlung photon spectrum.
Both spectra lead to similar plots as for LEP2 in Fig.3a.
The dominance near small $p_t$ of 
the beamstrahlung piece is due to the small final-state invariant 
mass preferred by the beamstrahlung spectrum.

\section{Conclusions}

We have argued in this letter that a measurement of the cross section
for heavy quark production in single-tag two-photon events is 
interesting, because it enables one to compare a well-controlled
QCD calculation with data at large $x$ and constrain the gluon content 
of the photon at small $x$.

Both in the $p_t$ and $y$ distributions very little distinguishes the NLO
results from the LO.  

Assuming a not overly pessimistic charm acceptance,
a measurement of the pointlike structure function seems feasible at
LEP2, and both the pointlike and hadronic structure functions are
certainly measurable at a future NLC.

\newpage

\begin{table}
\begin{center}
\begin{tabular}{||c|c|c|c|c||} \hline 
  $Q^2$       &  $x$                 & \multicolumn{3}{c||}{Events}  \\ \hline
  range       &  range  & $\mu_f = Q/2$ & $\mu_f = Q$ & $\mu_f = 2Q$ \\ \hline
  3.2 - 10    & $ 3.2 - 10.0 \cdot 10^{-4}$ & 3 & 2 & 2 \\ \cline{2-5}
              & $ 1.0 - 3.2 \cdot 10^{-3} $ & 25 & 21 & 20 \\ \cline{2-5}
              & $ 3.2 - 10.0 \cdot 10^{-3}$  & 106 & 94 & 88 \\ \cline{2-5}
              & $ 1.0 - 3.2 \cdot 10^{-2}$ & 339 & 314 & 296 \\ \cline{2-5}
              & $ 3.2 - 10.0 \cdot 10^{-2}$  & 919 & 884  & 857  \\ \cline{2-5}
              & $ 1.0 - 3.2 \cdot 10^{-1}$  & 2350 & 2220  & 2180  \\ \cline{2-5}
              & $ 3.2 - 10.0 \cdot 10^{-1}$  & 1040 & 916 & 845 \\ \hline
 10 - 32      & $ 1.0 - 3.2 \cdot 10^{-3}$ & 7 & 6 & 6 \\ \cline{2-5}
              & $ 3.2 - 10.0 \cdot 10^{-3}$  & 63 & 58  & 55  \\ \cline{2-5}
              & $ 1.0 - 3.2 \cdot 10^{-2}$  & 294 & 276  & 266  \\ \cline{2-5}
              & $ 3.2 - 10.0 \cdot 10^{-2}$  & 999 & 968  & 951  \\ \cline{2-5}
              & $ 1.0 - 3.2 \cdot 10^{-1}$  & 2910 & 2860 & 2830 \\ \cline{2-5}
              & $ 3.2 - 10.0 \cdot 10^{-1}$ & 3040 & 2880 & 2780 \\ \hline
 32 - 100     & $ 3.2 - 10.0 \cdot 10^{-3}$  & 6 & 6  & 5  \\ \cline{2-5}
              & $ 1.0 - 3.2 \cdot 10^{-2}$  & 54 & 51  & 50  \\ \cline{2-5}
              & $ 3.2 - 10.0 \cdot 10^{-2}$  & 248 & 241  & 235  \\ \cline{2-5}
              & $ 1.0 - 3.2 \cdot 10^{-1}$ & 825 & 814 & 808 \\ \cline{2-5}
              & $ 3.2 - 10.0 \cdot 10^{-1}$  & 1550 & 1520  & 1500  \\ \hline
 100 - 320    & $ 1.0 - 3.2 \cdot 10^{-2}$  & 5 & 5 & 5  \\ \cline{2-5}
              & $ 3.2 - 10.0 \cdot 10^{-2}$  & 45 & 44  & 43  \\ \cline{2-5}
              & $ 1.0 - 3.2 \cdot 10^{-1}$ & 196 & 194  & 193  \\ \cline{2-5}
              & $ 3.2 - 10.0 \cdot 10^{-1}$ & 511 & 509 & 508 \\ \hline
\end{tabular}
\end{center}
\caption{Number of events with charm quarks due to pointlike photons at LEP2 
in $e\gamma$ collisions.}
\end{table}

\newpage

\begin{table}
\begin{center}
\begin{tabular}{||c|c|c|c|c||} \hline 
 $Q^2$       &  $x$                 & \multicolumn{3}{c||}{Events}  \\ \hline
 range       &  range               & $\mu_f = Q/2$ & $\mu_f = Q$ & $\mu_f = 2Q$ \\ \hline
  3.2 - 10    & $ 3.2 - 10.0 \cdot 10^{-4}$ & 94 & 86 & 86 \\ \cline{2-5}
              & $ 1.0 - 3.2 \cdot 10^{-3} $ & 299 & 275 & 266 \\ \cline{2-5}
              & $ 3.2 - 10.0 \cdot 10^{-3}$  & 379 & 342 & 328 \\ \cline{2-5}
              & $ 1.0 - 3.2 \cdot 10^{-2}$ & 328 & 290 & 271 \\ \cline{2-5}
              & $ 3.2 - 10.0 \cdot 10^{-2}$  & 192 & 159  & 144  \\ \cline{2-5}
              & $ 1.0 - 3.2 \cdot 10^{-1}$  & 53 & 36  & 30  \\ \cline{2-5}
              & $ 3.2 - 10.0 \cdot 10^{-1}$  & 1 & 0 & 0 \\ \hline
 10 - 32      & $ 1.0 - 3.2 \cdot 10^{-3}$ & 121 & 116 & 116 \\ \cline{2-5}
              & $ 3.2 - 10.0 \cdot 10^{-3}$  & 333 & 320  & 313  \\ \cline{2-5}
              & $ 1.0 - 3.2 \cdot 10^{-2}$  & 410 & 385  & 372  \\ \cline{2-5}
              & $ 3.2 - 10.0 \cdot 10^{-2}$  & 306 & 279  & 267  \\ \cline{2-5}
              & $ 1.0 - 3.2 \cdot 10^{-1}$  & 111 & 95 & 88 \\ \cline{2-5}
              & $ 3.2 - 10.0 \cdot 10^{-1}$ & 5 & 3 & 3 \\ \hline
 32 - 100     & $ 3.2 - 10.0 \cdot 10^{-3}$  & 44 & 45  & 45  \\ \cline{2-5}
              & $ 1.0 - 3.2 \cdot 10^{-2}$  & 109 & 108  & 108  \\ \cline{2-5}
              & $ 3.2 - 10.0 \cdot 10^{-2}$  & 112 & 110  & 110  \\ \cline{2-5}
              & $ 1.0 - 3.2 \cdot 10^{-1}$ & 55 & 54 & 54 \\ \cline{2-5}
              & $ 3.2 - 10.0 \cdot 10^{-1}$  & 5  & 5  & 6  \\ \hline
 100 - 320    & $ 1.0 - 3.2 \cdot 10^{-2}$  & 11 & 11 & 12  \\ \cline{2-5}
              & $ 3.2 - 10.0 \cdot 10^{-2}$  & 23 & 24  & 24  \\ \cline{2-5}
              & $ 1.0 - 3.2 \cdot 10^{-1}$ & 16 & 17  & 18  \\ \cline{2-5}
              & $ 3.2 - 10.0 \cdot 10^{-1}$ & 3 & 3 & 3 \\ \hline
\end{tabular}
\end{center}
\caption{Number of events with charm quarks due to hadronic photons at LEP2 
in $e\gamma$ collisions.}
\end{table}

\newpage

\begin{table}
\begin{center}
\begin{tabular}{||c|c|c|c|c|c||} \hline 
  $Q^2$     &  $x$                 & \multicolumn{4}{c||}{Events}  \\ \hline
  range     &  range  & $c$(WW) & $b$(WW) & $c$(Beam) & $b$(Beam)\\ \hline
 32 - 100     & $ 1.0 - 3.2 \cdot 10^{-3}$  &1850&120&30&2  \\ \cline{2-6}
              & $ 3.2 - 10.0 \cdot 10^{-3}$  &2460&150&1170&70  \\ \cline{2-6}
              & $ 1.0 - 3.2 \cdot 10^{-2}$  &2420&130&4060&200   \\ \cline{2-6}
              & $ 3.2 - 10.0 \cdot 10^{-2}$  &2780&110&6510&260   \\ \cline{2-6}
              & $ 1.0 - 3.2 \cdot 10^{-1}$ &4470&150&9590&320   \\ \cline{2-6}
              & $ 3.2 - 10.0 \cdot 10^{-1}$  &6740&60&11,300&110   \\ \hline
 100 - 320    & $ 1.0 - 3.2 \cdot 10^{-3}$  &1040&80&0&0  \\ \cline{2-6}
              & $ 3.2 - 10.0 \cdot 10^{-3}$  &3790&300&220&20  \\ \cline{2-6}
              & $ 1.0 - 3.2 \cdot 10^{-2}$  &6230&430&3150&200  \\ \cline{2-6}
              & $ 3.2 - 10.0 \cdot 10^{-2}$  &10,300&510&12,200&580  \\ \cline{2-6}
              & $ 1.0 - 3.2 \cdot 10^{-1}$ &21,700&830&34,700&1320   \\ \cline{2-6}
              & $ 3.2 - 10.0 \cdot 10^{-1}$ &43,300&730&73,800&1240   \\ \hline
 320 - 1000   & $ 1.0 - 3.2 \cdot 10^{-2}$  &1050&110&70&10   \\ \cline{2-6}
              & $ 3.2 - 10.0 \cdot 10^{-2}$  &2310&160&1230&80   \\ \cline{2-6}
              & $ 1.0 - 3.2 \cdot 10^{-1}$  &5610&260&6790&310   \\ \cline{2-6}
              & $ 3.2 - 10.0 \cdot 10^{-1}$  &13,300&400&21,200&620    \\ \hline
1000 - 3200   & $ 1.0 - 3.2 \cdot 10^{-2}$  &70&10&0&0  \\ \cline{2-6}
              & $ 3.2 - 10.0 \cdot 10^{-2}$  &370&30&30&2   \\ \cline{2-6}
              & $ 1.0 - 3.2 \cdot 10^{-1}$  &1210&70&670&10    \\ \cline{2-6}
              & $ 3.2 - 10.0 \cdot 10^{-1}$  &3480&140&4200&160    \\ \hline
\end{tabular}
\end{center}
\caption{Number of events with charm and bottom quarks at a NLC 
in $e\gamma$ collisions.}
\end{table}

\newpage

\begin{table}
\begin{center}
\begin{tabular}{||c|c|c|c|c|c||} \hline 
Charm & $Q^2$ & $x$ & \multicolumn{3}{c||}{Events} \\ \cline{4-6}
Mass  & range & range & $\mu_f = Q/2$ & $\mu_f = Q$ &$\mu_f = 2Q$ \\ \hline
 1.3  &          &              & 2080 & 2140 & 2230 \\ \cline{4-6}
 1.5  & 32 - 100 &$ 1.0 - 3.2 \cdot 10^{-3}$ & 1880 & 1890 & 1930 \\ \cline{4-6}
 1.7  &          &          & 1740 & 1750 & 1790 \\ \hline
 1.3  &          &   & 346 & 356 & 370 \\ \cline{4-6}
 1.5  & 32 - 100 & $ 3.2 - 10.0 \cdot 10^{-4}$ & 317 & 322 & 329\\ \cline{4-6}
 1.7  &          &   & 296 & 305 & 313 \\ \hline
 1.3  &          &   & 1060 & 1100 & 1140 \\ \cline{4-6}
 1.5  & 100 - 320 & $ 1.0 - 3.2 \cdot 10^{-3}$  & 987 & 1024 & 1058 \\ \cline{4-6}
 1.7  &           &                             & 907 & 930 & 954 \\ \hline\hline
 1.3  &          &              & 3770 & 4080 & 4500 \\ \cline{4-6} 
 1.5  & 32 - 100 &$ 1.0 - 3.2 \cdot 10^{-3}$ & 3490 & 3740 & 4090 \\ \cline{4-6}
 1.7  &          &          & 3240 & 3480 & 3810 \\  \hline
 1.3  &          &   & 478 & 509 & 561 \\ \cline{4-6}
 1.5  & 32 - 100 & $ 3.2 - 10.0 \cdot 10^{-4}$ & 449 & 480 & 531 \\ \cline{4-6}
 1.7  &          &   & 419 & 452 & 502 \\ \hline
 1.3  &          &   & 829 & 916 & 1010 \\ \cline{4-6}
 1.5  & 100 - 320 & $ 1.0 - 3.2 \cdot 10^{-3}$  & 774 & 854 & 943 \\ \cline{4-6}
 1.7  &           &                             & 716 & 777 & 847 \\ \hline
\end{tabular}
\end{center}
\caption{
Number of events with charm at small $x$ at a NLC in $e\gamma$ collisions.
We used GRV {\protect \cite{GRV}} $\protect \overline {\rm MS}$
parton distributions for the top half and ACFGP {\protect \cite{acfgp}} ones 
for the bottom half.  }
\end{table}

\end{document}